\documentclass[longbibliography,aps,prb,reprint,superscriptaddress]{revtex4-1}
\usepackage[utf8]{inputenc}
\usepackage{graphicx}
\usepackage{hyperref}

\begin{document}

\title{Traces of Charge Density Waves in NbS$_2$}

\author{Maxime Leroux}
\altaffiliation{Present address: MPA-CMMS, Los Alamos National Laboratory, USA}
\affiliation{Univ. Grenoble Alpes, CNRS, Grenoble INP, Institut Néel, 38000 Grenoble France}

\author{Laurent Cario}
\affiliation{Institut des Matériaux de Nantes, Nantes, France}

\author{Alexei Bosak}
\affiliation{ESRF, The European Synchrotron, 71 Avenue des Martyrs, 38000 Grenoble, France}

\author{Pierre Rodière}
\email[Corresponding author]{}
\affiliation{Univ. Grenoble Alpes, CNRS, Grenoble INP, Institut Néel, 38000 Grenoble France}

\date{\today}

\begin{abstract}
Among Transition Metal Dichalcogenides (TMD), NbS$_2$ is often considered as the archetypal compound that does not have a Charge Density Wave (CDW) in any of its polytypes. By comparison, close iso-electronic compounds such as NbSe$_2$, TaS$_2$ and TaSe$_2$ all have CDW in at least one polytype.
Here, we report traces of CDW in the 2H polytype of NbS$_2$, using {diffuse x-ray scattering} measurements at 77\,K and room temperature. We observe 12 extremely weak satellite peaks located at $\pm$13.9$^\mathrm{o}$ from $\vec{a}^*$ and $\vec{b}^*$ around each Bragg peaks in the $(h,k,0)$ plane. These satellite peaks are commensurate with the lattice via $3\vec{q}-\vec{q}\,'=\vec{a}^*$, where $\vec{q}\,'$ is the 120$^{\circ}$ rotation of $\vec{q}$, and define two $\sqrt{13}\,a\times\sqrt{13}\,a$ superlattices in real space. These commensurate wavevectors and tilt angle are identical to those of the CDW observed in the 1T polytype of TaS$_2$ and TaSe$_2$. To understand this similarity and the faintness of the peaks, we discuss possible sources of local 1T polytype environment in bulk 2H-NbS$_2$ crystals.
\end{abstract}


\maketitle

\section{Introduction}

\begin{figure}
\includegraphics[width=\columnwidth]{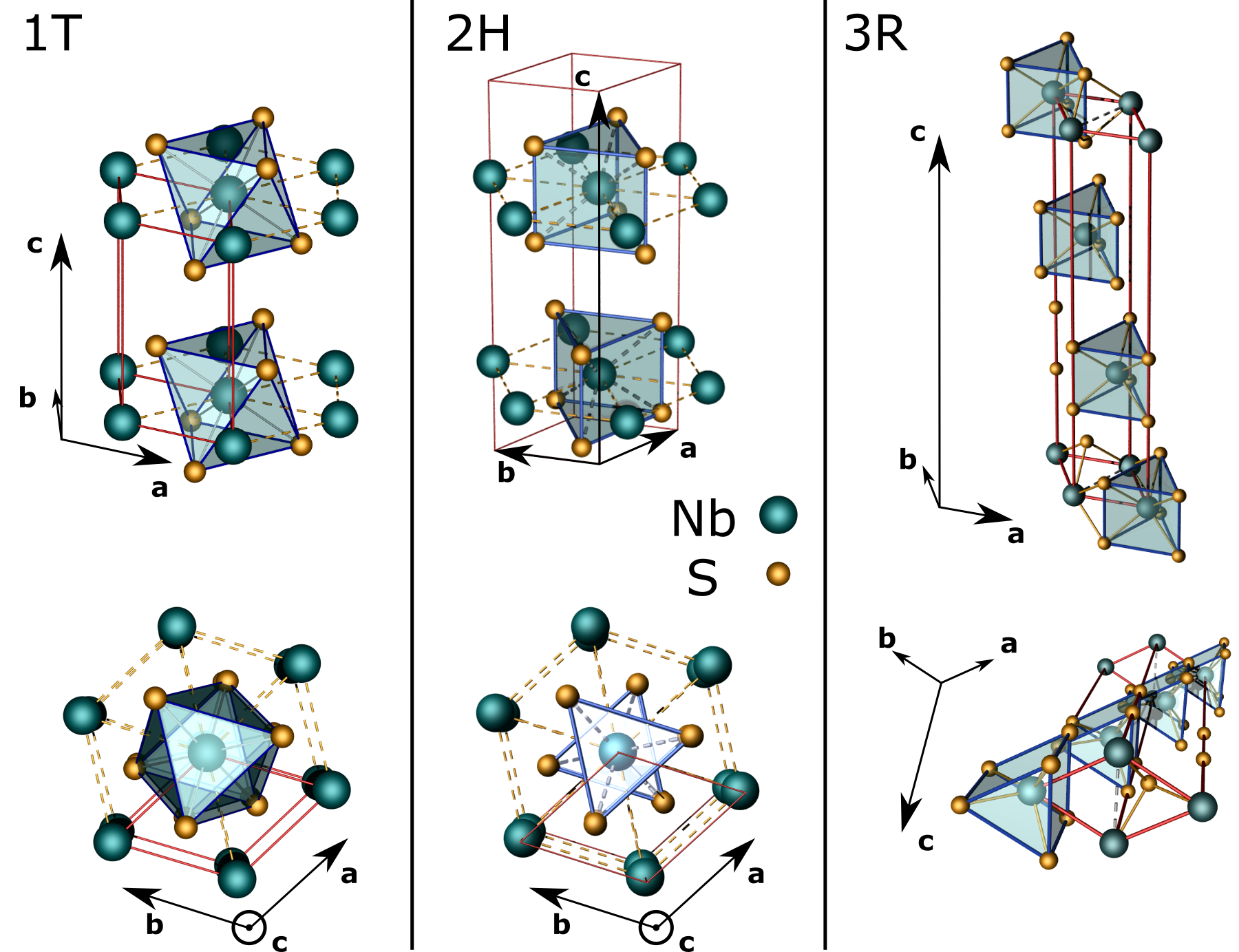}%
\caption{\label{fig:2H3R} \textbf{Cristallographic structures of the three known polytypes of NbS$_2$}: 1T reported only in thin film\cite{Carmalt2004_1st_1TNbS2synthesis} or monolayer form\cite{Chen20151TNbS2monolayerforH2}, and 2H and 3R found in bulk crystals form\cite{Fisher1980NbS2difficultsynthesis}. In the 1T polytype, the transition metal atom is in octahedral coordination and layers are stacked without rotation or in-plane translation. The 2H and 3R polytypes are composed of 1H single layers with the metal atom in trigonal prismatic coordination, but they differ by their stacking: rotation and no in-plane translation for 2H, in-plane translation and no rotation for 3R. The unit cell is indicated by solid red lines.}
\end{figure}

Transition Metal Dichalcogenides (TMD) have the generic formula MX$_2$, consisting of a transition metal M (Nb, Ta, Ti, Mo, W...) and a chalcogen X (S, Se, Te). They are layered materials with strong in-plane bonds and weak Van der Waals inter-layers interactions providing an important two dimensional (2D) character. The individual layers consist of a triangular lattice of transition metal atoms surrounded by chalcogens, and come in two forms named 1T and 1H. In 1T layers the transition metal atoms are surrounded by six chalcogens in octahedral (O$_h$) coordination, whereas in 1H layers the six chalcogens are in trigonal prismatic (D$_{3h}$) coordination. 
These two base layers have a wide range of possible stacking arrangements, called polytypes \cite{Wilson1975AdvPhysCDWTMDReview} (e.g. see Fig.~\ref{fig:2H3R}), which differ by the translation, rotation and ordering of the two base layers 1H and 1T. 
TMD polytypes are usually classified using Ramsdell's notation\cite{ramsdell1947studies}, which specifies the number of layers in the unit cell followed by a letter to indicate the lattice type and, when necessary, an additional alphanumeric character to distinguish between stacking sequences. Thus, a 1T polytype has 1 layer in a trigonal unit cell while a 2H polytype has 2 layers in a primitive hexagonal unit cell.
This distinction is especially important for TMD as polytypes of the same TMD compound can have dramatically different electronic properties spanning from semiconducting to metallic or superconducting\cite{Voiry2015TMD_polytypesynthesis_review}. 

TMD recently attracted renewed interest because their quasi 2D nature is similar to graphene and the tunability of their electronic properties is promising for novel electronic devices\cite{Wang2012TMDreviewnature,Vogel2015MRSBull}. In the case of metallic TMD, the 2D character and strong electron-phonon coupling makes them prone to electronic orderings such as Mott insulator or charge density waves (CDW) and superconductivity\cite{Wilson1975AdvPhysCDWTMDReview}. This multiplicity of possible ground states hold a great technological potential. For instance, a new \emph{orbitronics} concept has been proposed in TMD such as 1T-TaS$_2$, whereby switching between the orbital configurations and melting the CDW phase using ultrashort laser pulses would yield a complete and reversible semiconductor-to-metal transition\cite{Ritschel2015Orbitronics}. 

CDW are periodic modulations of the electronic density accompanied by a periodic distortion of the crystal lattice. CDW are usually caused either by a nesting vector of the Fermi surface inducing a peak in electronic susceptibility\cite{gruner}, or by a strong k-dependent electron-phonon coupling\cite{weber}. TMD are the first 2D compounds where CDW were observed\cite{Wilson1974PRL}, and TMD appear even more prone to CDW in single layer than in bulk form. For instance, in 2H-NbSe$_2$ the CDW transition temperature increases from 33\,K in bulk to 145\,K in monolayer form\cite{Xi2015highTcdwNbSe2,CalandraPRB2009monolayerNbSe2}.

However, among the metallic TMD, NbS$_2$ stands out as none of its polytypes have been reported to have a CDW. 
{In bulk form, only the 2H and 3R polytypes (trigonal prismatic coordination of Nb atoms) have been grown and CDWs have not been reported in either polytytpe\cite{FriendYoffe1987}}.
The trigonal prismatic coordination was found to be thermodynamically stable in bulk by DFT calculations\cite{LIU2014472}. The 1T polytype (octahedral coordination) has also been reported but only in single layer\cite{Chen20151TNbS2monolayerforH2} or thin film form\cite{Carmalt2004_1st_1TNbS2synthesis}, both also without CDW.

In the 2H polytype of NbS$_2$ that we study here, we previously showed that anharmonic effects prevent the formation of a CDW despite strong phonon modes softening\cite{Leroux2012anharmonicNbS2}. Thus, 2H-NbS$_2$ is just on the verge of a CDW and DFT calculations also hint at the proximity of density waves instabilities\cite{Leroux2012anharmonicNbS2, Guller2015DFTsdwNbS2, HeilPRL2017}.
The soft phonon modes do contribute to another electronic ordering: the metal-to-superconductor transition below $T_\mathrm{c}= 6$\,K\cite{Wilson1975AdvPhysCDWTMDReview}, in which they are the dominant contributor to anisotropic two-gap superconductivity\cite{Guillamon2008STMVortexcore,KACMARCIK2010S719,pribulova2010two,Diener2011TDONbS2,Leroux2012Hc1NbS2,HeilPRL2017}.
Yet, no other electronic phase has ever been found experimentally in 2H-NbS$_2$ using either: very pure crystals ($RRR=105$)\cite{Naito1982ResistivityNbS2}, low temperature (100\,mK)\cite{Guillamon2008STMVortexcore}, or high pressure\cite{JonesMorosinPRB1972pressureNbS2}. This is in contrast with the isoelectronic TMD 2H-NbSe$_2$ and 2H-TaS/Se$_2$ which all have CDW\cite{Wilson1975AdvPhysCDWTMDReview,Naito1982ResistivityNbS2}.
 
Here we find that there are faint traces of CDW in 2H-NbS$_2$ using {diffuse x-ray scattering}. This CDW wavevectors are the same as that of the commensurate CDW in 1T-TaS$_2$ and 1T-TaSe$_2$. Such 1T-like CDW has not been reported before for the NbS$_2$ compound. We suggest two mechanisms to explain both the symmetry and very small amplitude of the CDW we observe. Rotational stacking faults between 2H domains could be locally like a 1T-layer, or a very dilute amount of Nb in the van der Waals interlayer space could also present an octahedral coordination.


\section{Materials and methods}

Single crystals of 2H-NbS$_2$ were synthesized from an appropriate mixture of the elements that was sealed in an evacuated quartz tube. A large excess of sulfur was added to the mixture (20 \%) to act as a transporting agent and favor the formation of the 2H polytype. The tube was heated to 950$^\circ$C for 240\,h, slowly cooled down to 750$^\circ$C and subsequently quenched to room temperature. This synthesis yielded a powder containing single crystals with lateral sizes exceeding 200\,$\mu$m as shown in Ref.\cite{Diener2011TDONbS2}. Powder x-ray diffraction on several batches showed a predominance by volume of 99\% of 2H polytype (P6$_3/mmc$) versus 1\% of 3R (R$3m$), and the polytype of each single crystal used was checked individually using x-ray diffraction. We find lattice parameters $a = b = 3.33\,$\AA~and $c = 11.95\,$\AA. Superconducting properties and phonon spectrum of samples from this batch were published elsewhere\cite{Guillamon2008STMVortexcore,Kacmarcik2010CpNbS2,Diener2011TDONbS2,Leroux2012Hc1NbS2,Leroux2012anharmonicNbS2} and are in agreement with the literature\cite{Onabe1978ResistivityNbS2,Naito1982ResistivityNbS2}. Typical superconducting transition temperature is $T_\mathrm{c} = 6.05 \pm 0.4$\,K, as determined by AC specific heat\cite{Kacmarcik2010CpNbS2}. 

{Diffuse x-ray scattering} imaging was performed at beamline ID29 at the ESRF at a wavelength of 0.6966 \AA\ (17.798 keV) and using a PILATUS 6M detector 200\,mm away from the sample. 3600 pictures were acquired in three dimensions with 0.1$^\circ$ oscillations and 0.25\,s of exposure. Reconstruction of the $(h,k,0)$ plane was performed using CrysAlis software. Final reconstructions were made with locally developed software and Laue symmetry applied to remove the gaps between the detector elements. Inelastic x-ray scattering was performed at beamline ID28 at the ESRF using the Si (9,9,9) monochromator reflection giving an energy resolution of 2.6\,meV and a photon energy of 17.794 keV. Measurements were performed at 300 and 77\,K using a nitrogen cryostream cooler.

\section{Results}

\begin{figure}
\includegraphics[width=\columnwidth]{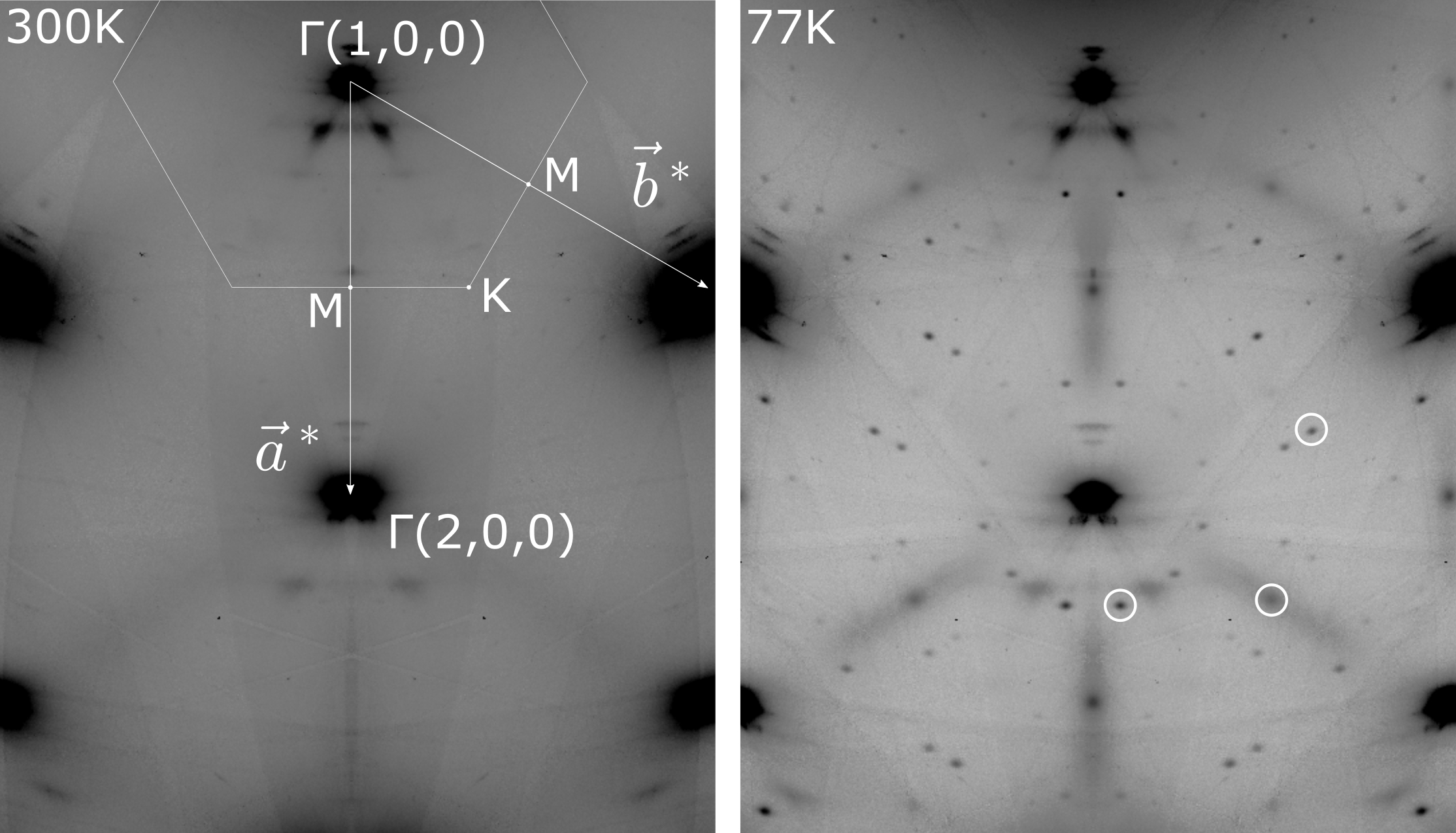}%
\caption{\label{fig:RT_77K}\textbf{{diffuse x-ray scattering} of 2H-NbS$_2$.} $(h,k,0)$ plane at 300\,K \textbf{(left)} and 77\,K \textbf{(right)} showing the hexagonal Brillouin zone, {the reciprocal space base vectors $\vec{a}^*$ and $\vec{b}^*$, the high symmetry points $\Gamma$, M and K, and the (1,0,0) and (2,0,0) Bragg peaks.} Elongated diffuse scattering is visible between Bragg peaks at 300\,K, and increases in intensity at 77\,K. It is caused by soft phonon modes and is not visible between each pair of Bragg peaks because of the longitudinal polarization of the soft phonon modes\cite{Leroux2012anharmonicNbS2}.
At 77\,K, three types of satellite peaks appear: a ring of twelve sharp peaks around each Bragg peak, a peak at the M point and 4 peaks around the M point. One example from each of these three sets of satellite peaks are indicated by white circles on the right panel}
\end{figure}

\begin{figure}
\includegraphics[width=\columnwidth]{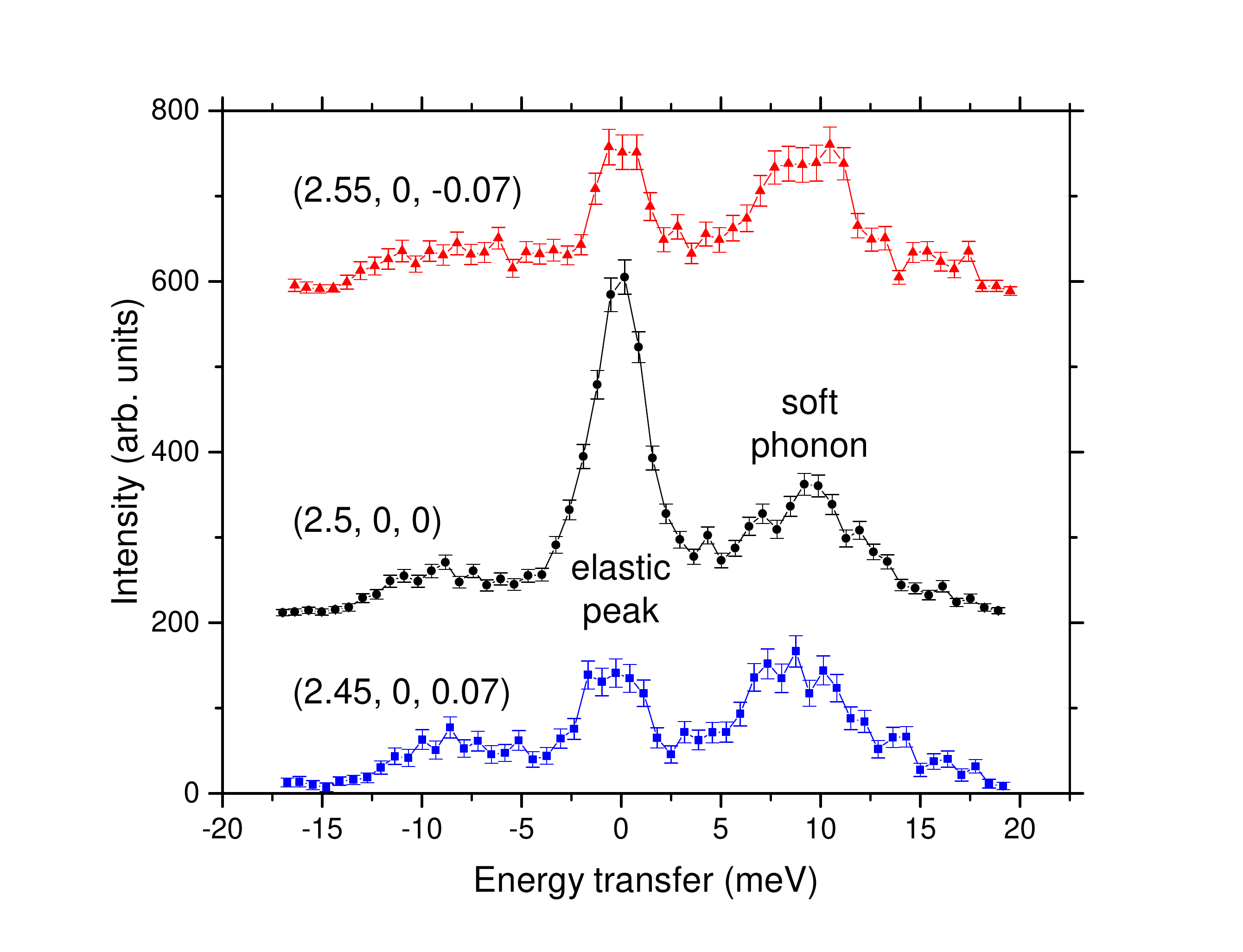}%
\caption{\label{fig:IXS}\textbf{Typical IXS spectra to determine the elastic or inelastic nature of the satellite peaks.} IXS energy spectra at 77\,K around the M point ($\vec{q} = (0.5,0,0)$) show that the elastic peak at zero energy transfer (static order) is the dominant contribution to the peak observed at the M point in diffuse scattering (energy integrated intensity). But the amplitude of this elastic peak is still comparable to that of a soft phonon. An elastic peak corresponds to a static diffracting object in real space.}
\end{figure}

Fig.~\ref{fig:RT_77K} shows the diffuse scattering in the $(h,k,0)$ plane reconstructed from diffuse scattering data, at 300 and 77\,K. The Bragg peaks amplitude is saturated on these images. The rocking curve of the (1,1,0) spot of a crystal from the same batch has a Full-Width at Half Maximum (FWHM) of 0.12$^\circ$ at room temperature, implying a Bragg peak FWHM of at most 0.0068\AA$^{-1}$ or 0.0036 $a^*$, i.e. an in-plane coherence length of at least 276 unit cells. 

At 300\,K, some diffuse scattering can be seen spanning the length between the different $\Gamma$M direction around each Bragg peak. This elongated diffuse scattering becomes salient at 77\,K.
It is caused by the broad softening of phonon modes around 1/3 of $a^*$ (2/3 of $\Gamma$M)\cite{Leroux2012anharmonicNbS2}. 

At 77\,K, Fig.~\ref{fig:RT_77K} also shows three types of satellite peaks: a peak at the M point, four peaks around the M point, and a ring of twelve peaks around each Bragg peak.
Inelastic x-ray scattering results, presented in Fig.~\ref{fig:IXS}, show that these peaks are all of an elastic nature, i.e. reflections of a static order, but with an amplitude similar to that of the soft phonon modes. {Comparison to the (1,1,0) Bragg peak, shows that these peaks are 5 orders of magnitude less intense}. Such a low intensity indicates that these peaks correspond either to very small atomic displacements or to displacements taking place in a very small fraction of the crystal.

\subsection{Satellite peak at the M point}

\textit{Ab initio} calculations\cite{Leroux2012anharmonicNbS2} find that the satellite peak at the M point corresponds to a maximum in electronic susceptibility. Considering its phonon-like amplitude, the peak at the M point could correspond to Friedel’s oscillations around impurities. However the peak FWHM is 0.036\,$a^*$, as shown in the lower panel of Fig.~\ref{fig:satpeakwidth}, which corresponds to a coherence length of about 30 unit cells. It therefore seems equally likely that the peak at the M point could correspond to an extremely faint CDW with a periodicity of $2\,a$ induced by the maximum in electronic susceptibility. Such faint $2\,a$ super-lattice spots have also been reported in 2H-NbSe$_2$\cite{CHEN1984}.

\begin{figure}
\includegraphics[width=\columnwidth]{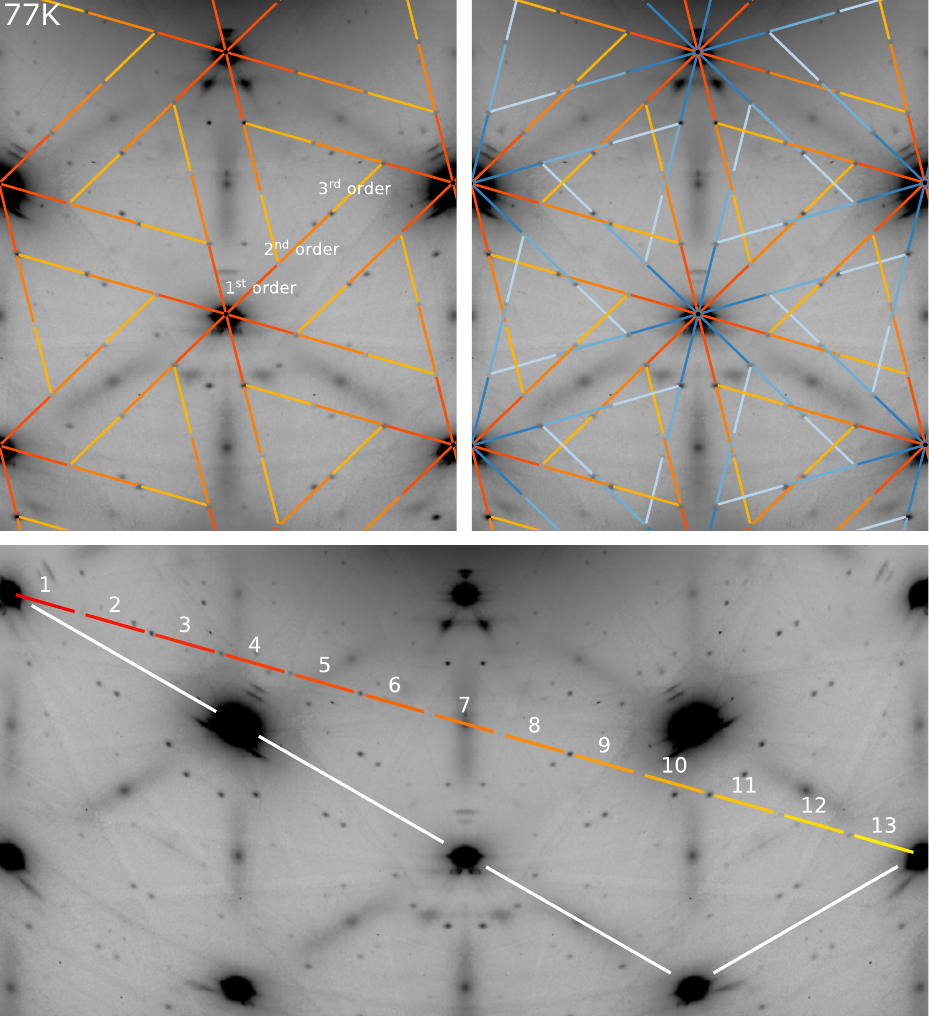}%
\caption{\label{fig:qvector}\textbf{Two interwoven commensurate superlattices} Diffuse scattering of 2H-NbS$_2$ in the $(h,k,0)$ plane at 77\,K.
The ring of twelve satellite peaks around each Bragg peak, and the four peaks around the M point can be indexed with two wavevectors. 
\textbf{(Upper left panel)} Subset of peaks indexed by $\vec{q_1} = \frac{3}{13}\,\vec{a}^* + \frac{1}{13}\,\vec{b}^*$, with $1^{\mathrm{st}}$, $2^{\mathrm{nd}}$ and $3^{\mathrm{rd}}$ order reflections.
\textbf{(Upper right panel)} Both subsets of peaks indexed by $\vec{q_1}$ in shades of red, and its mirror image $\vec{q_2} = \frac{4}{13}\,\vec{a}^* - \frac{1}{13}\,\vec{b}^*$ in shades of blue.
\textbf{(Lower panel)} $\vec{q_1}$, and $\vec{q_2}$ are commensurate with the crystal lattice via $13\,\vec{q_1} = 3\,\vec{a}^* + \vec{b}^*$. The wavevectors length is $||\vec{q_{1,2}}|| =\frac{1}{\sqrt{13}}||\vec{a}^*||$ so that each defines a $\sqrt{13}\,a\times\sqrt{13}\,a$ superlattice in real space. This is also geometrically equivalent to $3\vec{q_1}-\vec{q_1}'=\vec{a}^*$, where $\vec{q_1}'$ is $\vec{q_1}$ rotated by $+120^\circ$, which clearly appears in the upper panels.
{Note that the upper panels show the same region as in Fig.~\ref{fig:RT_77K}, and that the lower panel is an extended view centered on this same region.}
}
\end{figure}

\subsection{Other satellite peaks}

The ring of twelve satellite peaks around each Bragg peak, and the four peaks around the M point can all be indexed with only two wavevectors. The left panel of Fig.~\ref{fig:qvector} shows the wavevector: $\vec{q_1} = \frac{3}{13}\,\vec{a}^* + \frac{1}{13}\,\vec{b}^* \approx 0.231\,\vec{a}^* + 0.077\,\vec{b}^* $ and its $1^{\mathrm{st}}$, $2^{\mathrm{nd}}$ and $3^{\mathrm{rd}}$ order reflections. This wavevector corresponds to a deviation angle of $\arctan\left(\frac{\sqrt{3}}{7}\right)\approx13.9^{\circ}$ from $\vec{a}^*$. The right panel of Fig.~\ref{fig:qvector} shows both $\vec{q_1}$ in shades of red, and $\vec{q_2} = \frac{4}{13}\,\vec{a}^* - \frac{1}{13}\,\vec{b}^*  \approx 0.308\,\vec{a}^* - 0.077\,\vec{b}^* $ in shades of blue.

These two wavevectors are mirror image of each other, with length $||\vec{q_{1,2}}|| =\frac{1}{\sqrt{13}}||\vec{a}^*||$. They thus correspond to two commensurate $\sqrt{13}\,a\times\sqrt{13}\,a$ superlattices in real space.
Note that the commensurate relation $13\,\vec{q_1} = 3\,\vec{a}^* + \vec{b}^*$ is geometrically equivalent to $3\vec{q_1}-\vec{q_1}'=\vec{a}^*$, where $\vec{q_1}'$ is $\vec{q_1}$ rotated by $+120^\circ$. This clearly appears in the upper panels of Fig.~\ref{fig:qvector} where the $3^{\mathrm{rd}}$ order reflections from one Bragg peak coincide with the $1^{\mathrm{st}}$ order reflections from another. 

\begin{figure}
\includegraphics[width=\columnwidth]{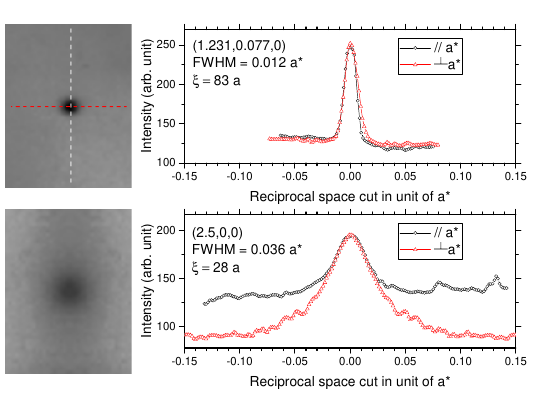}%
\caption{\label{fig:satpeakwidth}\textbf{Width of the satellite peaks} \textbf{(Upper panel)} cross-sections of the satellite peak at $(1,0,0)+\vec{q_1}$, which is part of the ring of 12 peaks around each Bragg peak. \textbf{(Lower panel)} cross-sections of a satellite peak at the M point. Note that the difference between the cross-section along and perpendicularly to $a^*$ are due to the background of diffuse scattering (caused by soft phonons) which disappears rapidly perpendicularly to $a^*$.}
\end{figure}

The presence of high order reflections evidences the long range coherence associated to these peaks or the non-sinusoidal character of the atomic displacements. 
The long range coherence is also evidenced by the small width of the peaks as shown in the upper panel of Fig.~\ref{fig:satpeakwidth}. The FWHM along $a^*$ of the satellite peak at (1.231, 0.077, 0) is 0.012 $a^*$ corresponding to a coherence length of $\approx 83$ unit cells. {These sharp peaks along a* correspond to rods of scattering along c* with FWHM of $\approx 0.5$\,c* i.e. 2 unit cells along the c-axis.}

The ring of twelve satellite peaks also has an intensity that follows the same extinction pattern as the elongated diffuse scattering, suggesting that it corresponds to a static longitudinal modulation. Indeed, the very specific angles at which the diffuse scattering is extinguished show that the underlying soft phonons (which cause the diffuse scattering) are polarized longitudinally with in-plane niobium displacements\cite{Leroux2012anharmonicNbS2}.
In more details, the scattered intensity depends on phonons polarization via the dynamical structure factor $G(Q,m)$~\cite{Burkel_IXS}
\begin{equation}
G(Q,m)=\left|\sum_{j}^{\text{unit cell}} f_j(\vec{Q}).\mathrm{e}^{-W_j}\left[\vec{Q}.\vec{\epsilon_j}(\vec{Q},m) \right] \sqrt{M_j} e^{i \vec{Q}.\vec{r_j}}\right|^2
\label{eqn:polarisation_phonon}
\end{equation}
where $f_j(\vec{Q})$ is the atomic form factor of atom $j$ at $\vec{r_j}$ with mass $M_j$; $\vec{\epsilon_j}(\vec{Q},m)$ is the unit displacement vector of atom $j$ in the $m$ phonon branch for a phonon wavevector $\vec{Q}$; and $\mathrm{e}^{-W_j}$ is the Debye-Waller factor of atom $j$.
Because $\vec{Q}.\vec{\epsilon_j}(\vec{Q},m)$ is zero for a phonon polarization perpendicular to $\vec{Q}$ (see Fig.11 in Ref.~\citenum{Burkel_IXS}), these extinctions indicate that the soft phonons are longitudinally polarized.
{We cannot distinguish the respective contribution of sulfur and niobium atoms to the longitudinal soft phonon modes in our data. However, the total scattered intensity is dominated by the contribution from niobium atoms as the mass and atomic form factor of niobium are larger than that of sulfur. This suggests that the displacements of the niobium atoms involved in the soft phonons would also be mostly longitudinal, i.e. in the ab plane.}

The commensurate wavevectors $\vec{q_1}$ and $\vec{q_2}$ are the same as those of the low temperature commensurate CDW in 1T-TaS$_2$\cite{ScrubyPhilMag1975} (semiconducting 1T$_3$ phase) and 1T-TaSe$_2$\cite{McMillanPRB1975,Wilson1974PRL}.
It is worth noting that this CDW is dominated by in-plane longitudinal displacements of Ta atoms in 1T-TaS$_2$\cite{ScrubyPhilMag1975}. Also, only one set of 6 peaks around each Bragg peaks is observed in the Ta based TMD. {But here we observe that both sets of 6 peaks are equivalently present, evidencing two sets of triple-q CDW, most likely from twinning in the crystal.}

We therefore conclude that the ring of peaks we observe in 2H-NbS$_2$ is the trace of a faint longitudinal periodic lattice distortion, appearing between 77 and 300\,K, and corresponding to two commensurate $\sqrt{13}\,a\times\sqrt{13}\,a$ CDW identical to that found in 1T Ta based TMD. 
In addition, as the commensurate CDW becomes incommensurate above 473\,K in 1T-TaSe$_2$ and 190\,K in 1T-TaS$_2$, this suggests the possibility of an incommensurate CDW in our crystal as well. As we observed no incommensurate peaks at 300\,K, this incommensurate CDW would have to occur in a temperature range between 77 and 300\,K.

Interestingly, in 1T-TaSe$_2$, the thrice degenerate wavevector of the high temperature incommensurate CDW becomes commensurate with the lattice by a rotation of 13.9$^{\circ}$ because it is not close enough to $1/3\,a^*$. {Indeed, according to Landau theory of CDW in TMD\cite{McMillanPRB1975}, this commensuration by rotation is a feature of the 1T polytype, whereas in the 2H polytype the CDW locks in with 1/3 of $a^*$\cite{McMillanPRB1975} (2/3 of $\Gamma$M). While we cannot preclude that the CDW we observe is a bulk phenomenon native to the 2H polytype, this would be the first $\sqrt{13}\,a\times\sqrt{13}\,a$ in a 2H TMD to our knowledge. In addition, the very short coherence length of the CDW peaks along c* supports the picture of a CDW occurring almost independently in each layer of the crystal. Therefore, we also consider the possibility that this CDW originates from a local 1T-like environment in a 2H crystal and we now discuss the possible origins of such environment.}

\section{Discussion}
In TMD, the dominance of trigonal prismatic (1H) or octahedral (1T) coordination can be classified by transition metal atoms. There are three typical cases. In the first case, as with titanium (Ti), the coordination is generally octahedral so that the dominant polytype is 1T. In the second case, such as niobium (Nb), the coordination is usually trigonal prismatic so that 2H or 3R polytypes are favored. In the third case, as with tantalum (Ta), both coordinations have similar energies, in which case various polytypes can be synthesized: 1T, 2H and mixed stackings of 1T and 1H layers such as 4Hb.\cite{FriendYoffe1987}

In NbS$_2$, the 3R polytype is the thermodynamically stable phase at room temperature\cite{Fisher1980NbS2difficultsynthesis}. The 2H polytype can also be synthesized at room temperature by quenching from $\approx1000$\,K. As for the 1T polytype of NbS$_2$, it has never been synthesized in bulk crystal form, but it can be stabilized by strain in thin film\cite{Carmalt2004_1st_1TNbS2synthesis} or monolayer\cite{Chen20151TNbS2monolayerforH2} forms.

Looking at the 2H structure in Fig.~\ref{fig:2H3R}, we emphasize that it has two possible rotational positions for each 1H layer, separated by a 60$^\circ$ rotation around the c-axis. This rotational position alternates between each 1H layers, so that, once an origin is given, the rotational positions of all 1H layers are fixed in an ideal crystal. In real crystals of the 2H structure, especially if synthesized by quenching, this opens up the possibility of rotational domains, where each domain has a different origin of the rotational positions. 

Most interestingly, at the junction of two rotational domains, there should be two 1H layers in the same rotational position stacked one onto the other (i.e. a locally 1H polytype), where the sulfur atoms are facing each other. This locally 1H polytype seems a priori unstable because of the geometrical repulsion between sulfur atoms in adjacent layers. In fact, such stacking of sulfur atoms does not occur in either of the three known polytypes 1T, 2H, or 3R and there are no known purely 1H polytype of NbS$_2$. Energetically, it seems much more likely that one of the sulfur atoms layer will move such that the sulfur atoms of one layer face the center of a triangle of sulfur atoms in the other layer. This reduces the geometrical repulsion, which brings the layers closer together and increases the orbital overlaps and van der Waals interactions. There are, however, several ways to displace the sulfur atoms layer.

One way, which does not involve changing the coordination of the Nb atoms, is for one of the 1H layer to slide by $(\frac{1}{3}, \frac{2}{3}, 0)$ or $(\frac{2}{3}, \frac{1}{3}, 0)$, yielding a locally 3R structure (which is non-centrosymmetric, hence the two possible sliding vectors). Such 3R-like stacking faults have actually been studied before in 2H-NbS$_2$. The study\cite{Katzke2002} concluded to the presence of 15\% of 3R-like stacking faults in powder samples of 2H-NbS$_2$ {(i.e. any two adjacent layers have a 15\% chance of having a faulty stacking)}.
We performed a similar analysis in our sample and found the presence of 18\% of 3R-like stacking faults.\cite{lerouxHALThesis}

A second way the sulfur atoms layer can move to reduce geometrical repulsion at the junction between domains, is a rotation by 60$^\circ$ around the c-axis. This changes the coordination of the Nb atom from trigonal prismatic to octahedral, and yields a single purely 1T layer. To some extent, this is similar to thin films and monolayer of 1T-NbS$_2$, where 1T layers are stabilized by strain at interfaces. This single 1T layer is only three-fold symmetric and can occur in two types which are mirror image of each other (or, equivalently, rotated by 60$^\circ$). The junctions between domains would yield both types equiprobably. This would naturally explain the presence of both wavevectors $\vec{q_1}$ and $\vec{q_2}$ yielding two $\sqrt{13}\,a\times\sqrt{13}\,a$ superlattices, instead of only one in pure 1T-TaS$_2$ and 1T-TaSe$_2$.

To our knowledge, this type of 1T-like stacking faults has not been studied before. In fact, considering that it involves a change of coordination of the Nb atom, a 1T-like stacking fault seems more energetic than the 3R-like stacking fault considered above. We can therefore expect that the 1T-like stacking fault occurs less frequently than the 3R-like ones. Yet, if the 1T-like CDW we observed in x-ray occurs only on such rare 1T-like stacking fault, it would explain why the CDW x-ray peaks are so faint.

Finally, another explanation for the presence of local 1T-like environment could be based on the presence of small clusters of extra Nb atoms intercalated in the van der Waals gap between layers. Indeed, Meerschaut and Deudon\cite{Meershault01} have reported that the 3R-NbS$_2$ phase is favored by an overdoping of Nb. This extra Nb is placed in the van der Waals gap between two layer of Nb in a trigonal prismatic coordination. Locally the Nb atom is surrounded by 6 chalcogen atoms in a octahedral coordination. Because of the Nb-Nb repulsion, this extra Nb atom is slightly shifted from the center of the octahedron \cite{Meershault01}. Thus, in our NbS$_2$ crystal, a local 1T-like environment could be associated to a small amount of extra Nb with a local octahedral coordination lying in the 3R-like stacking fault.

\section{Conclusions}
Using {diffuse x-ray scattering} in 2H-NbS$_2$, we observed very weak superlattice peaks corresponding to two longitudinal commensurate $\sqrt{13}\,a\times\sqrt{13}\,a$ periodic lattice distortion, identical to that associated with the CDW of 1T-TaSe$_2$ and 1T-TaS$_2$. Around each Bragg peaks in the $(h,k,0)$ plane, we found a series of 12 satellite peaks at $\pm$ 13.9$^{\circ}$ from $\vec{a}^*$ and $\vec{b}^*$, commensurate with the lattice through $3\vec{q_1}-\vec{q_1}'=\vec{a}^*$ or, equivalently, $13\,\vec{q_1} = 3\,\vec{a}^* + \vec{b}^*$. Inelastic x-ray scattering (IXS) measurements confirmed the predominantly elastic nature of these satellite peaks, but the amplitude of these peaks is almost as faint as that of soft phonons.
To our knowledge, no CDW has been reported in any polytypes of NbS$_2$.
We suggest that rotational disorder in the stacking of 1H layers, induces 3R-like stacking fault and, less frequently, single 1T layers at the interface between 2H rotational domains. Such rare and dilute 1T layers might be the support of this faint 1T-like CDW. A very dilute amount of Nb in the van der Waals interlayer space of 3R-like stacking fault could also present a 1T-like octahedral coordination.
\section{Acknowledgments}
M.L. acknowledges S. Eley for fruitful discussions.
This work was supported by the Neel Institute CNRS - University of Grenoble. We acknowledge the European Synchrotron Radiation Facility for provision of synchrotron radiation facilities. The experiments were performed on beamline ID28 and ID29.

\section*{References}

\bibliography{biblio}

\end{document}